\documentclass[12pt,draft]{iopart}
\usepackage{amsfonts}
\usepackage[english]{babel}
\usepackage{euscript}
\newcommand{\textfrac}[2]{{\textstyle\frac{#1}{#2}}}

\begin{document}

\title[Auto-B\"acklund Transformation for r-mmdKP equation]{%
Auto-B\"acklund Transformation for the r-th Double
\\
Modified Dispersionless Kadomtsev--Petviashvili Equation
\footnote[7]{The work of the first author was partially supported by the joint grant 09-01-92438-KE\_a of RFBR (Russia) and Consortium E.I.N.S.T.E.IN (Italy). The work of the second author was partially supported by the
RFBR grant 08-01-00054-à and by the grant of
Presidium of RAS "Fundamental Problems of Nonlinear Dynamics"}
}

\author{Oleg I. Morozov~$^\dag$ and Maxim V. Pavlov~$^\ddag$}

\address{$^\dag$~
Department of Mathematics, Moscow State Technical University
\\
of Civil Aviation, Kronshtadtskiy Blvd 20, Moscow 125993, Russia
\\
oim{\symbol{64}}foxcub.org}

\address{$^\ddag$~
Sector of Mathematical Physics, P.N. Lebedev Physical Institute
\\
of RAS, 53 Leninskii Ave.,  Moscow 119991, Russia
\\
M.V.Pavlov{\symbol{64}}lboro.ac.uk}

\begin{abstract}
We find auto-B\"acklund trans\-for\-ma\-ti\-on for the r-th double
modified dis\-per\-si\-on\-less Kadomtsev--Petviashvili equation.
\end{abstract}


\vskip 40 pt

The aim of this paper is to construct an auto-B\"acklund transformation for the r-th double modified dispersionless Kadomtsev--Petviashvili equation
\begin{eqnarray}
u_{yy} &=& u_{tx}
+ \left(\frac{(\kappa+1)\,u_y^2}{u_x^2}-\frac{u_t}{u_x}+\kappa\,u_x^\kappa u_y
+\frac{(\kappa+1)^2}{2\,\kappa+3}\,u_x^{2(\kappa+1)}
\right)\,u_{xx}
\nonumber
\\
&&
-\kappa\,\left(\frac{u_y}{u_x}+u_x^{\kappa+1}\right)\,u_{xy}
\label{rmmdKP}
\end{eqnarray}
with $\kappa \not \in \{-2, -3/2, -1\}$. This equation appears from the differential covering,
\cite{KV84,KLV,KV89},
\begin{eqnarray}
u_t &=& \left(
\frac{(\kappa+2)^2}{2\,\kappa+3}\,u_x^{2(\kappa+1)}
+(\kappa+2)\,w_x\,u_x^{\kappa+1}+\frac{\kappa+1}{2} \,w_x^2-w_y\right) u_x,
\label{covering_over_rmdKP_1}
\\
u_y &=& -\left(u_x^{\kappa+1}+w_x\right) u_x
\label{covering_over_rmdKP_2}
\end{eqnarray}
over the r-th modified dispersionless Kadomtsev--Petviashvili equation, \cite{Blaszak},
\begin{equation}
w_{yy} = w_{tx}+\left(\textfrac{1}{2}\,(\kappa+1)\,w_x^2+w_y\right)\,w_{xx} +\kappa\,w_x\,w_{xy},
\label{rmdKP}
\end{equation}
see
\cite{ChangTu},
\cite{KonopelchenkoAlonso},
\cite{Pavlov2006},
\cite{Morozov2010}.
Namely, excluding $w$ from (\ref{covering_over_rmdKP_1}) and (\ref{covering_over_rmdKP_2})  yields Eq. (\ref{rmmdKP}).

Using the method of \cite{Morozov2009} we obtain the following

\vskip 7 pt
\noindent
{\sc theorem} : {\it Equations
\begin{eqnarray}
v_t &=&
\frac{(\kappa+2)^2}{2\kappa+3}\,v_x^{2\kappa+3}-
(\kappa+2)\,\left(\frac{u_y}{u_x}+u_x^{\kappa+1}\right)\,v_x^{\kappa+2}
\nonumber
\\
&&
+\left(
\frac{u_t}{u_x}+(\kappa+2)\,u_x^\kappa u_y+\frac{(\kappa+1)(\kappa+2)}{2\kappa+3}
u_x^{2\kappa+2}
\right)\,v_x,
\label{covering_over_rmmdKP_1}
\\
v_y &=& -v_x^{\kappa+2} +\left(\frac{u_y}{u_x}+u_x^{\kappa+1}\right)\,v_x
\label{covering_over_rmmdKP_2}
\end{eqnarray}
define a covering for Eq. (\ref{rmmdKP}).}
\vskip 7 pt

The covering (\ref{covering_over_rmmdKP_1}), (\ref{covering_over_rmmdKP_2}) has the following remarkable property --- expressing $u_t$ and $u_y$ from (\ref{covering_over_rmmdKP_1}) and (\ref{covering_over_rmmdKP_2}) gives the same result as the change of variables $u\mapsto v$,
$v\mapsto u$. In other words, from  (\ref{covering_over_rmmdKP_1}) and (\ref{covering_over_rmmdKP_2}) it follows that $v$ satisfies to Eq. (\ref{rmmdKP}). Therefore
Eqs. (\ref{covering_over_rmmdKP_1}) and (\ref{covering_over_rmmdKP_2}) define an auto-B\"acklund transformation for Eq. (\ref{rmmdKP}).

\end{document}